\documentclass[useAMS,usenatbib,fleqn]{mn2e}

\def\apj{{ApJ}}                 
\def\apjl{{ApJ}}                
\def\apjs{{ApJS}}               

\def\mnras{{MNRAS}}             
\def\nar{{New A Rev.}}          
\def\prd{{Phys.~Rev.~D}}        
\usepackage{mathptmx}
\usepackage{amsmath}

\usepackage{epsf}
\usepackage{graphicx}

\newcommand{\mpc}{\rm {h^{-1}Mpc }}
\newcommand{\ud}{\textrm{d}}

\title[On the anisotropic density distribution]{On the anisotropic density distribution on large scales}

\author[P. Papai \& R. K. Sheth]{P\'eter P\'apai$^{1}$, Ravi K. Sheth$^{2,3}$\\ 
$^1$ Department of Physics, Faculty of Science, Prince of Songkla University, 15 Karnjanavanit Road, Hat Yai, Songkhla 90110, Thailand\\
$^2$ The Abdus Salam International Center for Theoretical Physics,
  Strada Costiera, 11, Trieste 34151, Italy\\
 $^3$ Center for Particle Cosmology, University of Pennsylvania, 
      209 S. 33rd St., Philadelphia, PA 19104, USA
}

\begin{document}
\pagerange{\pageref{firstpage}--\pageref{lastpage}}

\maketitle
\label{firstpage}

\begin{abstract}
Motivated by the recent detection of an enhanced clustering signal along the major axis of haloes in N-body simulations, we derive a formula for the anisotropic density distribution around haloes and voids on large scales. Our model, which assumes linear theory and that the formation and orientation of nonlinear structures are strongly correlated with the Lagrangian shear, is in good agreement with measurements. We also show that the measured amplitude is inconsistent with a model in which the alignment is produced by the initial inertia rather than shear tensor.

\end{abstract}


\section{INTRODUCTION}

The clustering of matter at late times provides important constraints on cosmological models. Our understanding of the signal is best on large scales, where it can be described by perturbation theory well \citep[see][]{Peebles}.  E.g., the Baryonic Acoustic Oscillations in the power spectrum (BAO), which appear as a spike in the two-point correlation function, lie in this regime \citep[see][and references to it]{Eisenstein2005}.  In addition to the simple correlation function, there are other ways to extract information from the matter distribution.  In redshift-space distorted measurements, the two-point correlation function is anisotropic (see \citealt{Kaiser1987} or the recent work of \citealt{Schlagenhaufer2012}), and this anisotropy can be used to constrain cosmological parameters.  However, certain real-space measures of clustering are also expected to be anisotropic.  Galaxy clusters are typically triaxial, and this triaxiality has long been known to align with the surrounding large scale structure \citep[e.g.,][and references therein]{Smargon2012}.  On smaller mass scales, galaxy spins are also known to align with the environment \citep[e.g.,][]{Lee2001,Zhang2009,Jones2010}.  Similar correlations have also been seen in simulations of voids \citep{Platen2008}.

Recently, \cite{Faltenbacher2012} showed that, in their numerical simulations of hierarchical gravitational clustering, the cross-correlation function between haloes and the surrounding mass was anisotropic:  this correlation between halo shapes and large scale structure extended even to the large scales relevant to BAO studies, and affected the zero-crossing of the correlation function.  This motivates our work, which attempts to model this anisotropy.  

Our model, which is described in Section~2, is based on the assumption that halo shapes \citep{Lee2005, Rossi2011} and orientations \citep{Lee2000,Lee2001} at late times are correlated with the properties of the initial Lagrangian field from which they formed.  This is a fundamental ingredient in models where haloes form from a triaxial collapse \citep{BM1996, SMT2001}.  In such models, the Lagrangian deformation or shear tensor plays a key role, as its eigenvalues can be used to distinguish between haloes, filaments, walls and voids.  We illustrate our model for the two extreme cases: haloes and voids. A final section discusses potential applications and extensions of our work.  

Although our analysis is general, when we illustrate our results, we will assume a $\Lambda$CDM model with $h = 0.73$, $\Omega_{\rm cdm}=0.205$, $\Omega_{bar}=0.045$, $\Omega_{\Lambda} = 0.75$, and $\sigma_8=0.9$.  These values allow a direct comparison with the simulations of \cite{Faltenbacher2012}, which we provide.

\section{ANISOTROPIC DENSITY DISTRIBUTION AROUND HALOES}

Despite the fact that haloes are highly nonlinear objects, their formation encodes information about the initial (Lagrangian) fields from which they formed \citep[e.g.,][]{PS, SMT2001}.  So, for example, one expects the shape, spatial orientation and spin of a halo to be correlated with the initial tidal field \citep[e.g.,][]{Lee2001,Rossi2011}, although nonlinear evolution may alter the form of this correlation \citep{vanHaarlem1993}.

\subsection{The shear}

The initial tidal field or shear tensor is defined as
\begin{equation}
 \xi_{ij}(\mathbf{q}) \equiv \frac{1}{\sigma_0} 
  \frac{\partial^2 \Phi (\mathbf{q})}{\partial q_i \partial q_j},
  \qquad {\rm where}\quad 
 \Delta \Phi(\mathbf{q}) \equiv \delta(\mathbf{q}),
 \label{shear}
\end{equation} 
where $\mathbf{q}$ is the Lagrangian spatial coordinate, $\Phi$ is the Lagrangian potential at $\mathbf{q}$, and $\sigma_0^2$ is the variance of the Lagrangian density $\delta$ smoothed on scale $R$.  This variance depends on the power spectrum $P(k)$ and the smoothing filter $W_R$:  
\begin{equation}
 \sigma_j^2 = \frac{1}{2\pi^2}\int dk\,k^{2(j+1)}\,P(k)\,W^2_R(k).
\end{equation}
For the $\Lambda$CDM parameters given earlier, we obtain the linear theory $P(k)$ from \emph{CAMB} \citep{CAMB}.  The resulting $\sigma_j$ decreases monotonically as $R$ increases.  As a result, in the excursion set description of haloes or voids \citep{Bond1991, Sheth2002}, $\sigma_0$ decreases as the halo mass or void radius increases.  This will be important when we wish to relate our results to the halo-based measurements in simulations.  

In what follows, we would like to estimate the anisotropy in the correlation between the density distribution at one position ($\mathbf{r}$) given that the shear at another position (which we will take to be the origin) satisfies some set of constraints.  That is to say, we are interested in 
\begin{eqnarray}
 \big<\delta(\mathbf{r})| C  \big> & = & 
   \frac{\int_{ C } \ud \pmb{\xi} \: p(\pmb{\xi}) \int \ud \delta(\mathbf{r})\, \delta(\mathbf{r})\, p(\delta(\mathbf{r})|\pmb{\xi})}
      {\int_{ C }\ud \pmb{\xi} \: p(\pmb{\xi})} ,\nonumber\\
     &=& 
      \frac{\int_{ C } \ud \pmb{\xi} \: p(\pmb{\xi})\: \big< \delta(\mathbf{r})|\pmb{\xi}\big>}{P( C )},
 \label{cond}
\end{eqnarray}
where $\pmb{\xi} = (\xi_{11},\xi_{22},\xi_{33},\xi_{12},\xi_{23},\xi_{13})$ is a 6-dimensional vector made from the components of the symmetric shear tensor and, for simplicity, we omit the distance argument if the quantity in question is taken at the origin: e.g. $\pmb{\xi} \equiv \pmb{\xi}(0)$.  In the expression above, $ C $ is the region in $\pmb{\xi}$-space where the conditions on the shear field (associated with halo or void formation) are satisfied; we use $P( C )$ to denote the integral over this region. For Gaussian initial conditions, $p(\pmb{\xi})$ is a multivariate Gaussian; in the principal axis frame, this distribution, first derived by \cite{Doroshkevich1970}, is given by our equation~(\ref{pl1l2l3}).  

\subsection{Average of $\delta(\mathbf{r})$ with conditions on the shear}

The Gaussianity of $p(\delta(\mathbf{r})|\pmb{\xi})$ means that 
\begin{equation}
 \big< \delta(\mathbf{r})|\pmb{\xi}\big> = 
    \big< \delta(\mathbf{r}) \otimes \pmb{\xi} \big>^{\top} \: \big< \pmb{\xi} \otimes \pmb{\xi} \big>^{-1}\: \pmb{\xi}
\end{equation}
where neither
 $\big< \delta(\mathbf{r})\, \otimes\, \pmb{\xi} \big>$ 
nor
 $\big< \pmb{\xi} \, \otimes\, \pmb{\xi} \big>^{-1}$ 
depend on $\pmb{\xi}$ \citep[see Appendix D of][]{BBKS}.  Therefore 
\begin{equation}
 \big< \delta(\mathbf{r})| C \big> = 
    \big< \delta(\mathbf{r}) \otimes \pmb{\xi} \big>^{\top} \: \big< \pmb{\xi} \otimes \pmb{\xi} \big>^{-1}\: 
    \big< \pmb{\xi} | C \big>.
 \label{dXi}
\end{equation}
The first two terms depend only on the correlations between $\delta$ at one position and the shear tensor $\pmb{\xi}$ at another.  Such correlations have been computed before \citep{Doroshkevich1970, BBKS, Wey1996, Crittenden2001,Catelan2001,Desjacques2008b,Lavaux2010,Rossi2012}. Although these expressions can be worked out exactly for the $6\times 6$ covariance matrix associated with $\pmb{\xi}$,  it is simpler to work in the coordinate system which is aligned with the principal axes of the shear tensor ($\xi_{12}=\xi_{13}=\xi_{23}=0$). In the rest of the paper, the subscript $D$ refers to the diagonal components of the shear: $\pmb{\xi}_D=(\xi_{11},\xi_{22},\xi_{33})$.  In this case, we find that 
\begin{equation}
 \big< \pmb{\xi}_D \otimes \pmb{\xi}_D \big>^{-1} =
  \begin{pmatrix}
      6 & -3/2 & -3/2 \\
   -3/2 &   6  & -3/2 \\
   -3/2 & -3/2 & 6 \label{cv}.
  \end{pmatrix}
\end{equation}

Similarly, using the form for $\big< \xi_{ij}(\mathbf{r})\xi_{kl} \big>$ that is given in the Appendix of \cite{Desjacques2008b} combined with the fact that 
 $\delta/\sigma_0\equiv \sum_{i}\xi_{ii}$, 
where $\sigma_0$ was defined in equation~(\ref{shear}), 
we find 
\begin{equation}
 \big<\delta(\mathbf{r}) \otimes \pmb{\xi}\big>_{ij}/\sigma_0= -\Delta_2(r) \hat{r}_i \hat{r}_j 
                + \frac{1}{3}(\Delta_0(r)+\Delta_2(r))\delta_{ij},
 \label{deltaxi}
\end{equation}
with $\mathbf{\hat{r}}$ being a unit vector and  
\begin{equation}
\Delta_{n}(r) \equiv \frac{1}{2\pi^2\sigma_0^2}\int \ud k k^2 j_n(rk)W_{R}(k)P(k),
 \label{pow}
\end{equation}
where $j_n$ is a spherical Bessel function.

Since 
\begin{equation}
 \sigma_0^2\,\Delta_{2}(r) \equiv \overline{\xi}_{2pt}(r) - \xi_{2pt}(r)
 \label{D2xi2}
\end{equation}
where 
 $\xi_{2pt}(r) = \sigma_0^2\,\Delta_0(r)$ 
is the usual angle-averaged two-point correlation function, and 
\begin{equation}
 \overline{\xi}_{2pt}(r) \equiv \frac{3}{r^3}\int_0^r \ud \tilde{r}\,
                               \tilde{r}^2\, \xi_{2pt}(\tilde{r})
 \label{xibar}
\end{equation}
is its volume average, equation~(\ref{deltaxi}) can be cast into a more intuitive form:
\begin{eqnarray}
 \frac{\big< \delta(\mathbf{r}) \otimes \pmb{\xi}\big>_{ij}}{\sigma_0} = 
   \frac{\xi_{2pt}(r)-\overline{\xi}_{2pt}(r)}{\sigma_0^2}\, \hat{r}_i\hat{r}_j
   + \frac{\overline{\xi}_{2pt}(r)}{3\sigma_0^2}\,\delta_{ij}.
\end{eqnarray}
In this expression, one should think of $\overline{\xi}_{2pt}$ as the overdensity within $r$, and $\overline{\xi}_{2pt}-\xi_{2pt}$ as the difference between the overdensity within $r$ and that at $r$ itself.  

Inserting equations~(\ref{cv}) and (\ref{deltaxi}) in equation~(\ref{dXi}) and averaging over $\phi$ in a spherical coordinate system defined as $(\hat{r}_1,\hat{r}_2,\hat{r}_3) = (\cos \theta, \sin \theta \cos \phi, \sin \theta \sin \phi )$ yield
\begin{equation}
 \big< \delta(r,\mu)| C  \big> = \big< \delta| C  \big> \,\Delta_0(r)\,
 - 5\, \big< \ell| C  \big>\, \Delta_2(r)\,P_2(\mu),
 \label{main}
\end{equation}
where $\mu=\cos \theta$, $P_2(\mu) = (3\mu^2-1)/2$ is a Legendre polynomial,
\begin{eqnarray}
 \big< \delta| C  \big> &\equiv& \sigma_0\,\big< \xi_{11}+\xi_{22}+\xi_{33} | C \big>,
 \label{delta}\\
 \big< \ell| C  \big> &\equiv& \sigma_0\,\big< \xi_{11}-\frac{\xi_{22}+\xi_{33}}{2}| C \big>.
 \label{cross}
\end{eqnarray}

In this form, it is clear that the first term on the rhs of equation~(\ref{main}), $\big< \delta| C  \big> \,\xi_{2pt}(r)/\sigma_0^2$, is the spherical average of the full expression.  Therefore, the prefactor should be thought of as a `linear bias factor' 
\begin{equation}
 b_C \equiv \big< \delta| C \big>/\sigma_0^2
 \label{bias}
\end{equation}
coming from the constraints $ C $.  (We provide an explicit example of this in the next section.)  The angular dependence comes from the second term, which, in fact, quantifies the local anisotropy. The result is intuitive: for spherical objects ($\ell = 0$), the anisotropy on large scales also disappears; while larger local anisotropy predicts larger anisotropy on large scales.

\begin{figure*}
    \begin{minipage}{175mm}
    \centerline{
      \leavevmode
      \epsfxsize=205mm    
      \epsfbox{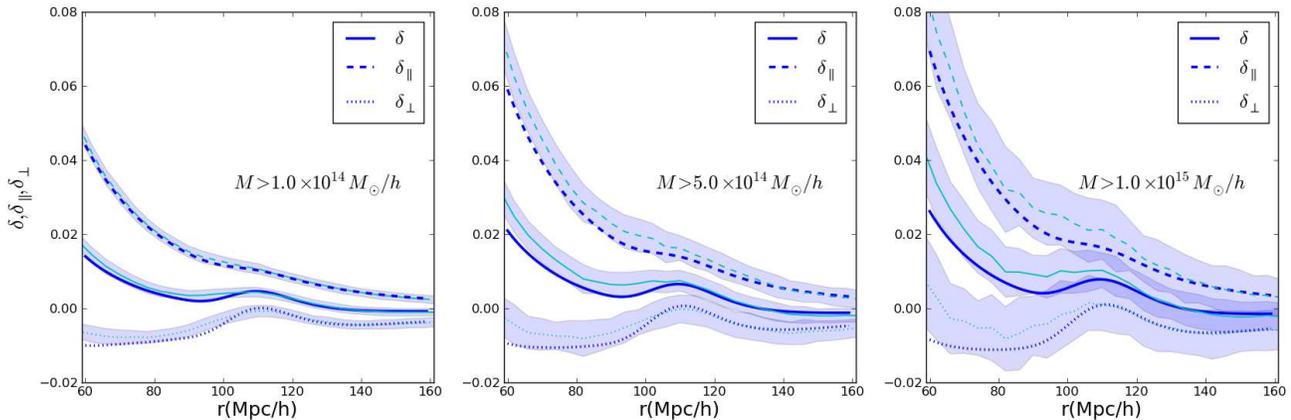}}
    	
        \caption{Comparison of our model with measurements in simulations, for a range of mass scales as indicated in each panel.  Thick blue curves show our equation~(\ref{main}), with $\sum_i \xi_{ii} >1.686/\sigma_0 $ and $\xi_{22,33}>\xi_{11}>0.41/\sigma_0$, and $\sigma_0$ determined by the mass.  Cyan shaded regions show data from Figure 1 of Faltenbacher et al. (2012).  In each case, theory and measurement have been averaged over the same range of orientation angles: $\cos \theta > 0.66$ and $\cos \theta < 0.33$ respectively.  
\label{fig:haloes}}
        \end{minipage}
\end{figure*}

Since both $b_C\,\xi_{2pt}(r)$ and its angle average $b_C\,\overline{\xi}_{2pt}(r)$ can be measured (indeed, these are the traditional 2-point measurements), our model can be written as 
\begin{eqnarray}
 \frac{\big< \delta(r,\mu)| C  \big>}{\big< \delta| C  \big> \,\Delta_0(r)}
  &=& 
  1 - A_C \, \frac{\Delta_2(r)}{\Delta_0(r)}\,P_2(\mu),
 \label{main0}
\end{eqnarray}
where 
\begin{equation}
 A_C \equiv 5\,\big< \ell| C  \big>/\big< \delta| C  \big>.
\end{equation}
The left hand side is the ratio of observables, and the right hand side shows that it is the product of a scale-independent amplitude, and a separable function of scale $r$ and angle.  Since the $r$ dependence is completely specified by measurable angle-averaged quantities, and the $\mu$ dependence is simply that of a quadrupole, the amplitude is the only free parameter in our model.  In this respect, the expression above should be thought of as providing a generic fitting formula with just one free parameter, the amplitude.  In our model, this amplitude encodes information about the alignment between the tracer field and the large scale environment.

\subsection{Illustrative constraints}
The averages over $ C $ in equations (\ref{delta}) and (\ref{cross}) can be calculated for many scenarios.  To gain intuition, suppose that we identify haloes with regions in the initial field for which all three eigenvalues were positive \citep[][]{Lee1998}.  Note that this is only realistic at large masses; a significant fraction of halos at lower masses has one negative eigenvalue \citep{Despali2012}.  In this case, requiring 
\begin{equation}
 C  = \{ \xi_{33}>\xi_{22}> \xi_{11}>0\} 
\end{equation}
means that 
\begin{eqnarray}
\big< \xi_{11}| C \big> &=& \frac{1}{30\sqrt{\pi}P( C )}\big( \sqrt{15}-\sqrt{10} \big), \\
\big< \xi_{22}| C \big> &=& \frac{1}{180\sqrt{\pi}P( C )}\big( 14\sqrt{15}-13\sqrt{10} \big) , \\
\big< \xi_{33}| C \big> &=& \frac{1}{180\sqrt{\pi}P( C )}\big( -5\sqrt{15} + 14\sqrt{10} \big) ,
\end{eqnarray}
where
\begin{equation}
P( C ) = \frac{1}{2} - \frac{\arctan (\sqrt{5})}{2\pi} - \frac{\sqrt{5}}{3\pi} \approx \frac{2}{25} \label{wtf}.
\end{equation}  
The above values were obtained from equations~(15), (A2), and (A3) of \cite{Lee1998}. (Our equation~(\ref{wtf}) is the integral of their equation~(15) for $\lambda_3>0$.  However, they state that this simple integral equals $2/25$ whereas it is in fact only a very good approximation to the exact answer, which we have given above.)  This makes
\begin{equation}
 b_C = \frac{\big< \delta| C \big>}{\sigma_0^2}
     = \frac{\sqrt{10}}{72\sqrt{\pi}P( C )}\frac{3\sqrt{6} - 2}{\sigma_0}
    \approx \frac{5}{3\sigma_0}
 \label{bpositive}
\end{equation}
and $A_C \approx -7/4$.  This shows that the bias factor increases as $\sigma_0$ decreases.  We remarked earlier that, in the excursion set approach, large masses have small $\sigma_0$.  Therefore, this model has the monopole part of the signal increasing as mass increases, but the ratio of the monopole to the quadrupole is independent of halo mass.

\subsection{More realistic conditions for haloes and voids}\label{conditions}

More realistic conditions for haloes and voids may require more than just the joint distribution of $\xi_{ii}$ in the principal axis frame (equation~\ref{pl1l2l3}).  E.g., the inertia tensor, and the alignment between the shear and inertia tensors may play a role.  But even in this simplest case, the moments of the distribution, that appear as $\big< \pmb{\xi}_D|C \big>$ in equation~(\ref{main}), can only be calculated analytically for the simplest $ C $ conditions.  Figure~\ref{fig:haloes} shows the result of evaluating equation~(\ref{main}) numerically with $ C $ given by the requirement that $\sum_i \xi_{ii} >1.686/\sigma_0$ and $\xi_{22,33}>\xi_{11}>0.41/\sigma_0$ for a range of choices of $\sigma_0$. These constraints on the $\xi_{ii}$  were motivated by the spherical collapse model and additional analysis in \cite{Lam2009}.  The values of $\sigma_0$ were chosen to match the halo masses quoted by \cite{Faltenbacher2012} in their analysis of the anisotropic clustering around haloes in simulations.

The agreement between theory and measurement is excellent for scales above $100 \mpc$ but it slightly underpredicts the clustering for lower scales. This discrepancy can be attributed to at least two reasons. First, our model of the Lagrangian patches which become haloes is crude, and may be inadequate. Second, our model is based on linear theory; \cite{vanHaarlem1993} argued that nonlinear effects matter, and more recent work has shown that nonlinear evolution will induce a quadrupole even if none is initially present, and will modify it if there is one initially \citep[although, on the scales of most interest here, this is expected to be subdominant]{Chan2012}. 


\begin{figure*}
    \begin{minipage}{175mm}
    \centerline{
      \leavevmode
      \epsfxsize=205mm    
      \epsfbox{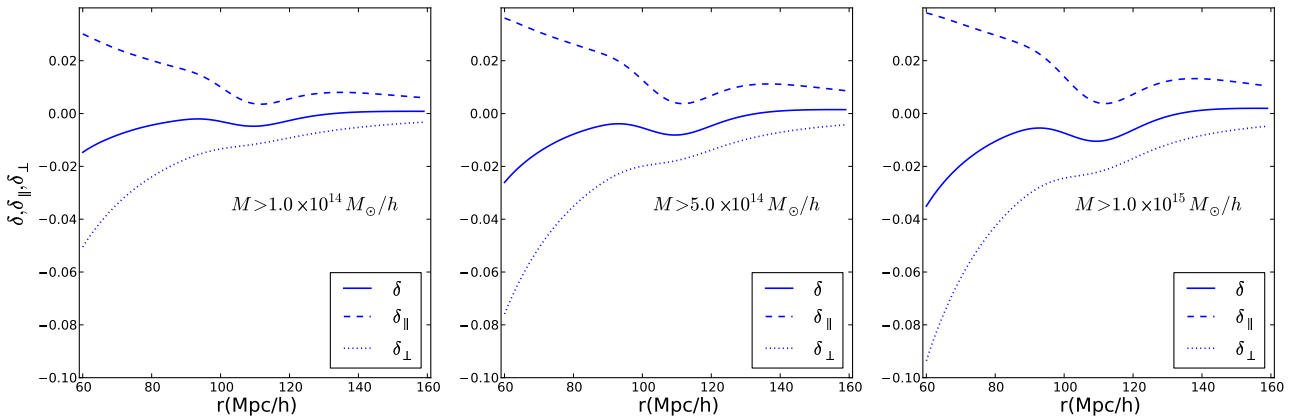}}	
    \caption{Equation~(\ref{main}) for voids: 
             $\sum_{i}\xi_{ii}<-2.8/\sigma_0$ and $0>\xi_{22,33}>\xi_{11}$. 
             \label{fig:voids}}
    \end{minipage}
\end{figure*}

Other nonlinear structures include filaments, sheets, and voids.  In triaxial models, this classification is related to the eigenvalues of the shear \citep{Shen2006}. All positive eigenvalues describe a halo, one negative gives a filament, two negatives a sheet,  and all three negative a void \citep[e.g.][]{Hahn2007}. For voids a reasonable set of conditions on the eigenvalues is $0>\xi_{22,33}>\xi_{11}$ and $\sum_i \xi_{ii} < -2.8/\sigma_0$ \citep[the latter condition comes from the spherical evolution model, e.g.,][]{Sheth2004}. Figure \ref{fig:voids} shows the result. In the direction of the major axis, there is a positive boost similarly to the case of haloes. Of course, since voids are driven towards spherical symmetry, their orientation may be harder to measure accurately.

\subsection{The shear vs the inertia tensor}\label{DvsI}

Haloes have been shown to align with the Lagrangian shear \citep{Dubinski1992,Lee2001}, and triaxial collapse models relate halo shapes to the initial shear \citep{SMT2001,Lee2005,Rossi2011}.   Similar arguments have been made for voids \citep{Platen2008}. These findings single out the shear tensor from others that could potentially define the axes of structures. An alternative to the shear is the initial inertia tensor (the matrix of second derivatives of the density field).  E.g., in the peaks model of \citep{BBKS}, haloes form at the peaks of the Lagrangian density field so their shape is given by the inertia tensor.  

The matter distribution around a peak with conditions on the inertia tensor is given by equation~(7.8) of \cite{BBKS}. Angle averaging their expression over $\phi$ yields
\begin{eqnarray}
\frac{\delta(r,\mu )}{\sigma_0} &=& \frac{\nu -\gamma x}{1-\gamma^2}\,\Delta_0(r) - \frac{\gamma \nu - x}{1-\gamma^2}\,\tilde{\Delta}_0(r) \nonumber\\
 && -5\big( \Lambda_1 - \frac{\Lambda_2 + \Lambda_3}{2} \big) \tilde{\Delta}_2(r) P_2(\mu ), \label{peak}
\end{eqnarray}
where $\Lambda_1,\Lambda_2$ and $\Lambda_3$ are the eigenvalues of
 $-\partial^2\delta(0)/\partial q_i \partial q_j$,
 $x = \sum_i \Lambda_i$, $\nu = \delta/\sigma_0$, 
 $\gamma = \sigma_1^2/(\sigma_0 \sigma_2)$, and 
\begin{equation}
 \tilde{\Delta}_{n}(r) \equiv \frac{1}{2\pi^2\sigma_0 \sigma_2}\int \ud k k^4 j_n(rk)W_{R}(k)P(k).
\end{equation}
It is worth noting that $\Lambda_1 - (\Lambda_2 + \Lambda_3)/2$ is $(3y+z)/2$, where $y$ and $z$ are the anisotropy parameters of \cite{BBKS}.  

\begin{figure*}
    \begin{minipage}{175mm}
    \centerline{
      \leavevmode
      \epsfxsize=205mm    
      \epsfbox{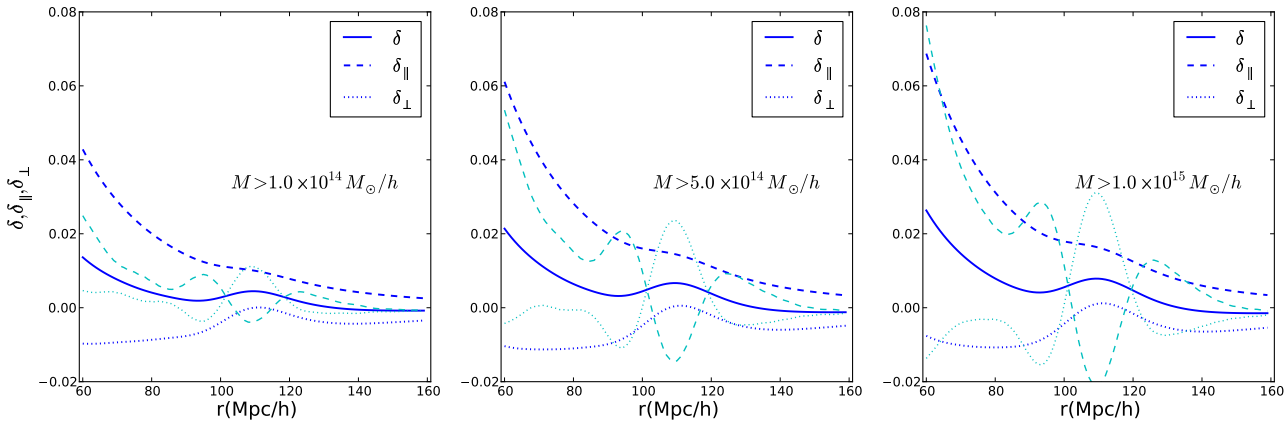}}
    \caption{Comparison of our model in which the deformation tensor plays a key role with the prediction of a model in which it is the inertia tensor which causes the correlation.  Blue curves show our model; they are the same as those in Figure~1.  Cyan curves show the sum of the spherical term of equation~(\ref{main}) with 100$\times$ the anisotropy term (the one coming from the final term of equation~\ref{peak}).  
             \label{fig:haloes2}}
    \end{minipage}
\end{figure*}

Although we have yet to average over the $\Lambda_i$, comparison with equation~(\ref{main}) shows that the first two terms in equation~(\ref{peak}) will yield the monopole, and the final term a quadrupole.  Although the quadrupole here depends on the eigenvalues $\Lambda$ of the inertia tensor in the same way that the quadrupole in equation~(\ref{main}) depends on the eigenvalues of the shear tensor, we might expect the amplitude here will be much smaller.  This is because the integral which defines $\tilde{\Delta}_2$ has two additional powers of $k$ compared to that which defines $\Delta_2$.  However, we must also check that the average over the $\Lambda_i$ does not yield a large amplitude to compensate for this difference.  

To see that this will not happen, note that on large scales, the leading order contribution to the monopole is given by the first term on the rhs of equation~(\ref{peak}).  With the peak constraints, the quantity $\tilde{\nu} \equiv (\nu-\gamma x)/(1-\gamma^2)$ is a Gaussian variate \citep[see][]{Wey1996,Lavaux2010,Rossi2012}, and it is independent of the $\Lambda_i$. This means that we can choose the constraints on $\tilde{\nu}$ to match the monopole of equation~(\ref{main}), leaving us to perform an independent average over the distribution of parameters $x$, $y$, and $z$ \cite[e.g. Appendix~C of][]{BBKS}.  The resulting angular dependence is multiplied by 100 in order to produce the cyan curves shown in Figure~\ref{fig:haloes2}. Since the nonlinear evolution effects described by \cite{Chan2012} cannot, on their own, account for the signal seen by \cite{Faltenbacher2012}, we conclude that the initial shear tensor matters for the angular dependence, whereas the inertia tensor does not.

\section{DISCUSSION}
\label{discussion}

In this paper, we calculated the anisotropy in the linear density field when conditions are placed on the Lagrangian shear field.  If the shear field is strongly correlated with the shapes and orientations of nonlinear haloes, then this calculation should be closely related to the anisotropy of the halo-mass cross-correlation function, which is most easily seen when the mass field around haloes is stacked after aligning along the major axis of the halo \citep[e.g.][]{Faltenbacher2012}.

For haloes, our model (equation~\ref{main}) captures the main features of the cross-correlation function measurement.  The signal along the long axis is stronger than perpendicular to it, but produces a less prominent BAO feature (Figure~\ref{fig:haloes}).  We predict a similar effect for voids (Figure \ref{fig:voids}).  Overall, the signal is slightly weaker on scales below $100 \mpc$ than in simulations. This may be due to inadequacies in our crude model which relates halo formation to the local shear; or nonlinear evolution may have had a small effect (see Section \ref{conditions}). 

Formally, the approximations involved can be summed up as 
\begin{equation}
P\big( \delta (\mathbf{r})|\mathbf{\hat{e}}_h \big) \approx P_{GS}\big( \delta(\mathbf{r})|\mathbf{\hat{e}}_{sh}=\mathbf{\hat{e}}_h, C  \big),
\end{equation}
where the lhs is the true distribution of $\delta$ at $\mathbf{r}$ relative to a halo oriented in the direction of $\mathbf{\hat{e}}_h$, while the rhs is the usual Gaussian conditional probability of $\delta$ with $ C $ denoting the previously discussed conditions on the shear tensor (Section~\ref{conditions}) and $\mathbf{\hat{e}}_{sh}$ the direction of the eigenvector that belongs to the smallest eigenvalue. Improvements can be devised along the following identity:
\begin{eqnarray}
P(\delta(\mathbf{r})|\mathbf{\hat{e}}_h) & = & \int  P\big( \delta(\mathbf{r})|\{ \mathbf{\hat{e}}_{sh},\xi_{ii} \} , \mathbf{\hat{e}}_{h} \big) \nonumber \\
& \times & P\big( \{ \mathbf{\hat{e}}_{sh},\xi_{ii} \} | \mathbf{\hat{e}}_{h} \big)\ud \{ \mathbf{\hat{e}}_{sh},\xi_{ii}\}, \label{appr}
\end{eqnarray}
where $\{ \mathbf{\hat{e}}_{sh},\xi_{ii}\}$ and $\ud\{ \mathbf{\hat{e}}_{sh},\xi_{ii}\}$ are a parametrization of the shear and its volume element respectively. As the final orientation of a halo ($\mathbf{\hat{e}}_{h}$) results from the nonlinear evolution of the Lagrangian field, it is a function of the local Lagrangian field and its derivatives. In Section~\ref{DvsI}, we showed that on large scales in the Gaussian limit higher order derivatives had a small effect on $\big< \delta(\mathbf{r})|\dots \big>$ compared to the shear. In this limit, the first term behind the integral in equation~(\ref{appr}) turns into $ P\big( \delta(\mathbf{r})|\{ \mathbf{\hat{e}}_{sh},\xi_{ii} \} , \mathbf{\hat{e}}_{h} \big) \approx P_{GS}\big( \delta(\mathbf{r})|\{ \mathbf{\hat{e}}_{sh},\xi_{ii} \}) $. To improve on this approximation, the nonlinear evolution of haloes has to be understood better. The second term in the integral is equally challenging. 

A practical approach can be taken by fitting a phenomenological formula  to measurements in N-body simulations, similarly to \cite{Lee2001}, who parametrized the angular momentum of a halo as a function of the shear. E.g., a simple model is given by 
\begin{eqnarray}
P\big( \{ \mathbf{\hat{e}}_{sh},\xi_{ii} \} | \mathbf{\hat{e}}_{h} \big) & \approx & \frac{P(\{ \xi_{ii} \} |  C  )}{2\pi} \times \nonumber \\
& & \bigg( P_{\parallel }\delta_{D}(1-\mathbf{\hat{e}}_{h}\cdot \mathbf{\hat{e}}_{sh})+(1-P_{\parallel}) \bigg),\label{ali} 
\end{eqnarray}
which assumes that the eigenvalues of the shear ($\xi_{ii}$) are independent of the orientation of the halo ($\mathbf{\hat{e}}_{h}$) and that the shear is perfectly aligned with haloes in $P_{\parallel}\times 100$ per cent of the time, otherwise its orientation is completely random.  With this, the spherical part of equation~(\ref{main}) remains the same, while the anisotropy gets multiplied by $P_{\parallel}$. Figure~\ref{fig:haloes} implies a strong correlation, so $P_{\parallel}$ must be close to $1$. This is a conjecture that can be verified by a direct measurement of the halo-shear alignment. In general, a more complex model of alignment would introduce higher order Legendre polynomials in the expansion of $\delta(r,\mu)$. 

The same argument holds for voids as well. As the ellipticity of voids is less prominent \citep{Sheth2004}, their orientation can be measured with a lower accuracy. In a model of alignments, this would increase the randomness. E.g. in equation~(\ref{ali}), $P_{\parallel}$ would be smaller thus reducing the measured anisotropy.     

We also argued that the measured amplitude is inconsistent with a model in which the alignment is produced by the initial inertia rather than shear tensor (Section~\ref{DvsI}).  

Absent a model for halo or void formation, our equation~(\ref{main}) may be treated as a one-parameter family which, given the spherically averaged measurement, describes the anisotropy.  This parameter depends only on the local shear, and so may be used to constrain models of halo formation and alignment.  Further work can be done to incorporate redshift distortions and nonlinearities into the model.  Also, tests on simulations are needed in order to identify systematics that can affect the validity of the model.  Finally, we are in the process of checking if this sort of measurement can yield useful constraints on modified gravity models.

\section{ACKNOWLEDGMENT}

We would like to thank A. Faltenbacher for providing his data in 
electronic format, and the anonymous referee for a conscientious review 
of our work. PP is grateful for a CEI fellowship.  
RKS was supported in part by NSF 0908241 and NASA NNX11A125G.  
He is grateful to B. Bassett for organizing a mini-workshop at AIMS 
in January 2012 where he had interesting discussions with A. Faltenbacher 
and U.-L. Pen about this effect, as well as the participants of the 
Cape Town Cosmology School 2012 for inspiration.  
He is also grateful to the group at LUTH in Meudon Observatory for their
hospitality during June 2012.
Thanks also to R. van de Weygaert for pointing us to helpful and relevant earlier work on this subject.

\appendix

\section{DETAILS OF COMPUTATION PRESENTED IN SECTION 2}

In this short Appendix, we write out the steps that lead to equations~(\ref{deltaxi}) and (\ref{main}).

To derive Equation~(\ref{deltaxi}), it is convenient to work in Fourier space: 
\begin{eqnarray}
\big<\delta(\mathbf{r}) \otimes \pmb{\xi}(\mathbf{r'})\big>_{ij} = 
\int \frac{\ud^3k \ud^3k'}{(2\pi)^6} e^{i\mathbf{k}\mathbf{r}}e^{-i\mathbf{k'}\mathbf{r'}} \big<\delta(\mathbf{k}) \otimes \pmb{\xi}^*(\mathbf{k'})\big>_{ij} .
\end{eqnarray} 
$\xi_{ij}$ translates into $-\hat{k}'_i\hat{k}'_j\delta(\mathbf{k'})$ in Fourier space. The integral over $\mathbf{k'}$ can be carried out easily as $\big< \delta(\mathbf{k}) \delta^*(\mathbf{k'}) \big> = (2\pi)^3 \delta_D\big( \mathbf{k}-\mathbf{k'} \big) P(k)$, also $\mathbf{r'}$ can be set to 0 for convenience. The easiest way to proceed is to work with spherical harmonics. Using the plane wave expansion
\begin{equation} 
e^{i\mathbf{k}\mathbf{r}} = 4\pi\sum_{l=0}^{\infty}\sum_{m=-l}^{l}i^lj_l(kr)Y_{lm}(\hat{\mathbf{r}})Y^*_{lm}(\hat{\mathbf{k}})
\end{equation}
along with a special spherical coordinate system, which is defined by the Cartesian coordinates and allows us to express the $\hat{k}_i\hat{k}_j$ type of terms with spherical harmonics (e.g. $\hat{k}_1^2 = \frac{4}{3}\sqrt{\frac{\pi}{5}}Y_{20}(\hat{\mathbf{k}})+\frac{2}{3}\sqrt{\pi}Y_{00}(\hat{\mathbf{k}})$, etc.), the angular part of the remaining integral $\int \ud^3k = \int \ud k k^2 \int \ud \hat{\mathbf{k}}$ can be carried out. The result of this tedious but straightforward computation is equation~(\ref{deltaxi}). 

As we are only interested in  
\begin{equation}
 \big< \delta(\mathbf{r})| C \big> = 
    \big< \delta(\mathbf{r}) \otimes \pmb{\xi}_D \big>^{\top} \: \big< \pmb{\xi}_D \otimes \pmb{\xi}_D \big>^{-1}\: 
    \big< \pmb{\xi}_D | C \big>,
\end{equation}
(take note of subscript $D$ denoting the diagonal terms of the shear), we only need to deal with the diagonal part of equation~(\ref{deltaxi}) in further calculations: 
\begin{eqnarray}
 \big< \delta(\mathbf{r})| C \big> &=& \bigg( \Delta_0(r) + \Delta_2(r)\big( 6\hat{r}_1^2-\frac{3}{2}(\hat{r}_2^2+\hat{r}^2_3)-1 \big) \bigg) \nonumber \\
& &\times \sigma_0\big< \xi_{11}|C \big> 
+ \mathrm{cyc}.
\end{eqnarray}
The angular average of this expression around the major axis of a halo (lets say axis 1) can be derived by adopting spherical coordinates$(\hat{r}_1,\hat{r}_2,\hat{r}_3) = (\cos \theta, \sin \theta \cos \phi, \sin \theta \sin \phi )$ and averaging over $\phi$. The result is equation~(\ref{main}).

Finally, the Doroshkevich formula \citep{Doroshkevich1970} is
\begin{eqnarray}
 p(\lambda_1,\lambda_2,\lambda_3) = \frac{3375}{8\sqrt{5}\pi\sigma^6}\exp \bigg( -\frac{3I_1^2}{\sigma^2} + \frac{15I_2}{2\sigma^2}\bigg)\nonumber\\
 \times (\lambda_1-\lambda_2)(\lambda_2-\lambda_3)(\lambda_1-\lambda_3),
 \label{pl1l2l3}
\end{eqnarray}
where $\lambda_i$ is the $i$th eigenvalue of the inertia tensor and $\sigma^2$ is the variance of the mass. $I_1 = \sum_i \lambda_i$ and $I_2=\lambda_1 \lambda_2 + \mathrm{cyc}$.

\label{lastpage}


\begin{thebibliography}{}

\bibitem[\protect\citeauthoryear{{Bardeen}, {Bond}, {Kaiser} \&
  {Szalay}}{{Bardeen} et~al.}{1986}]{BBKS}
{Bardeen} J.~M.,  {Bond} J.~R.,  {Kaiser} N.,    {Szalay} A.~S.,  1986, \apj,
  304, 15

\bibitem[\protect\citeauthoryear{{Bond}, {Cole}, {Efstathiou} \&
  {Kaiser}}{{Bond} et~al.}{1991}]{Bond1991}
{Bond} J.~R.,  {Cole} S.,  {Efstathiou} G.,    {Kaiser} N.,  1991, \apj, 379,
  440

\bibitem[\protect\citeauthoryear{{Bond} \& {Myers}}{{Bond} \&
  {Myers}}{1996}]{BM1996}
{Bond} J.~R.,  {Myers} S.~T.,  1996, \apjs, 103, 1

\bibitem[\protect\citeauthoryear{Catelan \& Porciani}{Catelan \&
  Porciani}{2001}]{Catelan2001}
Catelan P.,  Porciani C.,  2001, \mnras, 323, 713

\bibitem[\protect\citeauthoryear{{Chan}, {Scoccimarro} \& {Sheth}}{{Chan}
  et~al.}{2012}]{Chan2012}
{Chan} K.~C.,  {Scoccimarro} R.,    {Sheth} R.~K.,  2012, \prd, 85, 083509

\bibitem[\protect\citeauthoryear{Crittenden, Natarajan, Pen \&
  Theuns}{Crittenden et~al.}{2001}]{Crittenden2001}
Crittenden R.,  Natarajan P.,  Pen U.,    Theuns T.,  2001, \apj, 559, 552

\bibitem[\protect\citeauthoryear{{Desjacques}}{{Desjacques}}{2008}]{Desjacques2008b}
{Desjacques} V.,  2008, \mnras, 388, 638

\bibitem[\protect\citeauthoryear{Despali, Tormen \& Sheth}{Despali
  et~al.}{2012}]{Despali2012}
Despali G.,  Tormen G.,    Sheth R.~K.,  2012, \mnras, submitted

\bibitem[\protect\citeauthoryear{{Doroshkevich}}{{Doroshkevich}}{1970}]{Doroshkevich1970}
{Doroshkevich} A.~G.,  1970, Astrophysics, 6, 320

\bibitem[\protect\citeauthoryear{Dubinski}{Dubinski}{1992}]{Dubinski1992}
Dubinski J.,  1992, \apj, 401, 441

\bibitem[\protect\citeauthoryear{Eisenstein}{Eisenstein}{2005}]{Eisenstein2005}
Eisenstein D.,  2005, \nar, 49, 360

\bibitem[\protect\citeauthoryear{{Faltenbacher}, {Li} \& {Wang}}{{Faltenbacher}
  et~al.}{2012}]{Faltenbacher2012}
{Faltenbacher} A.,  {Li} C.,    {Wang} J.,  2012, \apjl, 751, L2

\bibitem[\protect\citeauthoryear{{Hahn}, {Carollo}, {Porciani} \&
  {Dekel}}{{Hahn} et~al.}{2007}]{Hahn2007}
{Hahn} O.,  {Carollo} C.~M.,  {Porciani} C.,    {Dekel} A.,  2007, \mnras, 381,
  41

\bibitem[\protect\citeauthoryear{{Jones}, {van de Weygaert} \&
  {Arag{\'o}n-Calvo}}{{Jones} et~al.}{2010}]{Jones2010}
{Jones} B.~J.~T.,  {van de Weygaert} R.,    {Arag{\'o}n-Calvo} M.~A.,  2010,
  \mnras, 408, 897

\bibitem[\protect\citeauthoryear{{Kaiser}}{{Kaiser}}{1987}]{Kaiser1987}
{Kaiser} N.,  1987, \mnras, 227, 1

\bibitem[\protect\citeauthoryear{{Lam}, {Sheth} \& {Desjacques}}{{Lam}
  et~al.}{2009}]{Lam2009}
{Lam} T.~Y.,  {Sheth} R.~K.,    {Desjacques} V.,  2009, \mnras, 399, 1482

\bibitem[\protect\citeauthoryear{{Lavaux} \& {Wandelt}}{{Lavaux} \&
  {Wandelt}}{2010}]{Lavaux2010}
{Lavaux} G.,  {Wandelt} B.~D.,  2010, \mnras, 403, 1392

\bibitem[\protect\citeauthoryear{{Lee}, {Jing} \& {Suto}}{{Lee}
  et~al.}{2005}]{Lee2005}
{Lee} J.,  {Jing} Y.~P.,    {Suto} Y.,  2005, \apj, 632, 706

\bibitem[\protect\citeauthoryear{Lee \& Pen}{Lee \& Pen}{2000}]{Lee2000}
Lee J.,  Pen U.,  2000, \apjl, 532, L5

\bibitem[\protect\citeauthoryear{{Lee} \& {Pen}}{{Lee} \&
  {Pen}}{2001}]{Lee2001}
{Lee} J.,  {Pen} U.-L.,  2001, \apj, 555, 106

\bibitem[\protect\citeauthoryear{{Lee} \& {Shandarin}}{{Lee} \&
  {Shandarin}}{1998}]{Lee1998}
{Lee} J.,  {Shandarin} S.~F.,  1998, \apj, 500, 14

\bibitem[\protect\citeauthoryear{Lewis, Challinor \& Lasenby}{Lewis
  et~al.}{2000}]{CAMB}
Lewis A.,  Challinor A.,    Lasenby A.,  2000, \apj, 538, 473

\bibitem[\protect\citeauthoryear{Peebles}{Peebles}{1980}]{Peebles}
Peebles P.,  1980, The large-scale structure of the universe.
Princeton Univ Press, Princeton, NJ

\bibitem[\protect\citeauthoryear{{Platen}, {van de Weygaert} \&
  {Jones}}{{Platen} et~al.}{2008}]{Platen2008}
{Platen} E.,  {van de Weygaert} R.,    {Jones} B.~J.~T.,  2008, \mnras, 387,
  128

\bibitem[\protect\citeauthoryear{{Press} \& {Schechter}}{{Press} \&
  {Schechter}}{1974}]{PS}
{Press} W.~H.,  {Schechter} P.,  1974, \apj, 187, 425

\bibitem[\protect\citeauthoryear{{Rossi}}{{Rossi}}{2012}]{Rossi2012}
{Rossi} G.,  2012, \mnras, 421, 296

\bibitem[\protect\citeauthoryear{{Rossi}, {Sheth} \& {Tormen}}{{Rossi}
  et~al.}{2011}]{Rossi2011}
{Rossi} G.,  {Sheth} R.~K.,    {Tormen} G.,  2011, \mnras, 416, 248

\bibitem[\protect\citeauthoryear{{Schlagenhaufer}, {Phleps} \&
  {S{\'a}nchez}}{{Schlagenhaufer} et~al.}{2012}]{Schlagenhaufer2012}
{Schlagenhaufer} H.~A.,  {Phleps} S.,    {S{\'a}nchez} A.~G.,  2012, \mnras,
  425, 2099

\bibitem[\protect\citeauthoryear{{Shen}, {Abel}, {Mo} \& {Sheth}}{{Shen}
  et~al.}{2006}]{Shen2006}
{Shen} J.,  {Abel} T.,  {Mo} H.~J.,    {Sheth} R.~K.,  2006, \apj, 645, 783

\bibitem[\protect\citeauthoryear{Sheth, Mo \& Tormen}{Sheth
  et~al.}{2001}]{SMT2001}
Sheth R.,  Mo H.,    Tormen G.,  2001, \mnras, 323, 1

\bibitem[\protect\citeauthoryear{{Sheth} \& {Tormen}}{{Sheth} \&
  {Tormen}}{2002}]{Sheth2002}
{Sheth} R.~K.,  {Tormen} G.,  2002, \mnras, 329, 61

\bibitem[\protect\citeauthoryear{{Sheth} \& {van de Weygaert}}{{Sheth} \& {van
  de Weygaert}}{2004}]{Sheth2004}
{Sheth} R.~K.,  {van de Weygaert} R.,  2004, \mnras, 350, 517

\bibitem[\protect\citeauthoryear{{Smargon}, {Mandelbaum}, {Bahcall} \&
  {Niederste-Ostholt}}{{Smargon} et~al.}{2012}]{Smargon2012}
{Smargon} A.,  {Mandelbaum} R.,  {Bahcall} N.,    {Niederste-Ostholt} M.,
  2012, \mnras, 423, 856

\bibitem[\protect\citeauthoryear{{van de Weygaert} \& {Bertschinger}}{{van de
  Weygaert} \& {Bertschinger}}{1996}]{Wey1996}
{van de Weygaert} R.,  {Bertschinger} E.,  1996, \mnras, 281, 84

\bibitem[\protect\citeauthoryear{{van Haarlem} \& {van de Weygaert}}{{van
  Haarlem} \& {van de Weygaert}}{1993}]{vanHaarlem1993}
{van Haarlem} M.,  {van de Weygaert} R.,  1993, \apj, 418, 544

\bibitem[\protect\citeauthoryear{Zhang, Yang, Faltenbacher, Springel, Lin \&
  Wang}{Zhang et~al.}{2009}]{Zhang2009}
Zhang Y.,  Yang X.,  Faltenbacher A.,  Springel V.,  Lin W.,    Wang H.,  2009,
  \apj, 706, 747

\end{thebibliography}
\end{document}